\documentclass[aps,prl,twocolumn,superscriptaddress,showpacs,preprintnumbers,amsmath,amssymb]{revtex4}

\usepackage{graphicx} 
\usepackage{dcolumn}  
\usepackage{subfigure}
\usepackage[dvips]{color}

\newcommand{\gev} {\ensuremath{\, {\mathrm{GeV}}     }}
\newcommand{\mev} {\ensuremath{\, {\mathrm{MeV}}     }}
\newcommand{\gevc}{\ensuremath{\, {\mathrm{GeV}/c^2} }}
\newcommand{\mevc}{\ensuremath{\, {\mathrm{MeV}/c^2} }}

\newcommand{\ecm} {\ensuremath{ E_{\mathrm{c.m.}} }}
\newcommand{\gisr}{\ensuremath{ \gamma_{ISR} }}

\newcommand{\RMS} {\ensuremath{ M^2_{\mathrm{rec}} }}

\newcommand{\ee}        {\ensuremath{ e^+e^- }}
\newcommand{\eedd}      {\ensuremath{ e^+e^- \to D   \overline D     }}

\newcommand{\eeddp}     {\ensuremath{ e^+e^- \to D^0 D^- \pi^+ }}
\newcommand{\eeddpg}    {\ensuremath{ e^+e^- \to D^0 D^- \pi^+ \gamma_{ISR} }}

\newcommand{\psiddt} {\ensuremath{\psi(4415)\to D \overline D{}^{*}_2(2460) }}

\newcommand{\ps}    {\ensuremath{ \psi(4415) }}
\newcommand{\ddp}   {\ensuremath{ D^0 D^- \pi^+ }}
\newcommand{\ddpf}  {\ensuremath{ D D \pi }}

\newcommand{\ddnbp}   {\ensuremath{ D^0 \overline D{}^0 \pi^+ }}
\newcommand{\dpdmp}   {\ensuremath{ D^+ D^- \pi^+ }}
\newcommand{\dstmdnp} {\ensuremath{ D^{*-} D^0 \pi^+ }}

\newcommand{\dt}    {\ensuremath{ \overline D{}^{*}_2(2460)   }}
\newcommand{\dtch}  {\ensuremath{ D{}^{*}_2(2460)^+ }}
\newcommand{\dtnb}  {\ensuremath{ \overline D{}^{*}_2(2460)^0 }}

\newcommand{\ddt}   {\ensuremath{ D  \overline D{}^{*}_2(2460)   }}
\newcommand{\ddtch} {\ensuremath{ D^- D{}^{*}_2(2460)^+ }}
\newcommand{\ddtnb} {\ensuremath{ D^0\overline D{}^{*}_2(2460)^0 }}

\newcommand{\mddp}  {\ensuremath{ M_{D^0 D^-\pi^+} }}

\newcommand{\dmpip} {\ensuremath{ D^- \pi^+ }}
\newcommand{\dnpip} {\ensuremath{ D^0 \pi^+ }}
\newcommand{\dpi}   {\ensuremath{ D   \pi^+ }}

\newcommand{\mdn}   {\ensuremath{ M_{D^0}   }}
\newcommand{\mdm}   {\ensuremath{ M_{D^-}   }}

\newcommand{\misid} {\ensuremath{ e^+e^- \to D^0 D^- \pi^+ \pi^0  }}

\newcommand{\eeddppg} {\ensuremath{ e^+e^- \to D^0    D^- \pi^+ \pi^0_{\mathrm{miss}} \gamma_{ISR} }}
\newcommand{\eeddpgg} {\ensuremath{ e^+e^- \to D^{*0} D^- \pi^+ \gamma_{ISR} }}
\newcommand{\eeison}  {\ensuremath{ e^+e^- \to D^0 \overline D{}^0 \pi^+ \pi^-_{\mathrm{miss}} \gamma_{ISR}}}
\newcommand{\eeisoch} {\ensuremath{ e^+e^- \to D^+  D^- \pi^+ \pi^-_{\mathrm{miss}} \gamma_{ISR} }}

\begin{document}

\title{\quad\\[0.5cm] Observation of \psiddt\ decay using initial-state radiation}


\affiliation{Budker Institute of Nuclear Physics, Novosibirsk}
\affiliation{University of Cincinnati, Cincinnati, Ohio 45221}
\affiliation{The Graduate University for Advanced Studies, Hayama}
\affiliation{Hanyang University, Seoul}
\affiliation{University of Hawaii, Honolulu, Hawaii 96822}
\affiliation{High Energy Accelerator Research Organization (KEK), Tsukuba}
\affiliation{Hiroshima Institute of Technology, Hiroshima}
\affiliation{Institute of High Energy Physics, Chinese Academy of Sciences, Beijing}
\affiliation{Institute of High Energy Physics, Vienna}
\affiliation{Institute of High Energy Physics, Protvino}
\affiliation{Institute for Theoretical and Experimental Physics, Moscow}
\affiliation{J. Stefan Institute, Ljubljana}
\affiliation{Kanagawa University, Yokohama}
\affiliation{Korea University, Seoul}
\affiliation{Kyungpook National University, Taegu}
\affiliation{\'Ecole Polytechnique F\'ed\'erale de Lausanne (EPFL), Lausanne}
\affiliation{University of Ljubljana, Ljubljana}
\affiliation{University of Maribor, Maribor}
\affiliation{University of Melbourne, School of Physics, Victoria 3010}
\affiliation{Nagoya University, Nagoya}
\affiliation{National Central University, Chung-li}
\affiliation{National United University, Miao Li}
\affiliation{Department of Physics, National Taiwan University, Taipei}
\affiliation{H. Niewodniczanski Institute of Nuclear Physics, Krakow}
\affiliation{Nippon Dental University, Niigata}
\affiliation{Niigata University, Niigata}
\affiliation{University of Nova Gorica, Nova Gorica}
\affiliation{Osaka City University, Osaka}
\affiliation{Osaka University, Osaka}
\affiliation{Panjab University, Chandigarh}
\affiliation{Princeton University, Princeton, New Jersey 08544}
\affiliation{Saga University, Saga}
\affiliation{University of Science and Technology of China, Hefei}
\affiliation{Seoul National University, Seoul}
\affiliation{Sungkyunkwan University, Suwon}
\affiliation{University of Sydney, Sydney, New South Wales}
\affiliation{Tata Institute of Fundamental Research, Mumbai}
\affiliation{Toho University, Funabashi}
\affiliation{Tohoku Gakuin University, Tagajo}
\affiliation{Tohoku University, Sendai}
\affiliation{Department of Physics, University of Tokyo, Tokyo}
\affiliation{Tokyo Institute of Technology, Tokyo}
\affiliation{Tokyo Metropolitan University, Tokyo}
\affiliation{Tokyo University of Agriculture and Technology, Tokyo}
\affiliation{Virginia Polytechnic Institute and State University, Blacksburg, Virginia 24061}
\affiliation{Yonsei University, Seoul}

 \author{G.~Pakhlova}\affiliation{Institute for Theoretical and Experimental Physics, Moscow} 
  \author{I.~Adachi}\affiliation{High Energy Accelerator Research Organization (KEK), Tsukuba} 
  \author{H.~Aihara}\affiliation{Department of Physics, University of Tokyo, Tokyo} 
  \author{K.~Arinstein}\affiliation{Budker Institute of Nuclear Physics, Novosibirsk} 
  \author{V.~Aulchenko}\affiliation{Budker Institute of Nuclear Physics, Novosibirsk} 
  \author{T.~Aushev}\affiliation{\'Ecole Polytechnique F\'ed\'erale de Lausanne (EPFL), Lausanne}\affiliation{Institute for Theoretical and Experimental Physics, Moscow} 
  \author{A.~M.~Bakich}\affiliation{University of Sydney, Sydney, New South Wales} 
  \author{V.~Balagura}\affiliation{Institute for Theoretical and Experimental Physics, Moscow} 
  \author{E.~Barberio}\affiliation{University of Melbourne, School of Physics, Victoria 3010} 
  \author{I.~Bedny}\affiliation{Budker Institute of Nuclear Physics, Novosibirsk} 
% ...
\collaboration{The Belle Collaboration}

\begin{abstract}
We report measurements of the exclusive cross section for $\ee\to\ddp$
over the center-of-mass energy range 4.0\gev\ to 5.0\gev\ with
initial-state radiation and the first observation of the decay
$\ps\to\ddp$. From a study of the resonant substructure in \ps\ decay
we conclude that the $\ps\to\ddp$ decay is dominated by \psiddt.  We
obtain $\mathcal{B}(\ps\to\ddp_{\mathrm
  {non-resonant}})/\mathcal{B}(\ps\to\ddt\to\ddp)<0.22$ at 90\% C.L.
The analysis is based on a data sample collected with the Belle
detector with an integrated luminosity of $673$ $\mathrm{fb}^{-1}$.
\end{abstract}


\maketitle
\setcounter{footnote}{0}

The \ps\ resonance, the heaviest well established $J^{PC}=1^{--}$
charmonium state, was first observed thirty years ago by the
MARK-I~\cite{mark:cs} and DASP~\cite{dasp:cs} collaborations.
Subsequently, although additional \ee\ annihilation cross section
measurements in the region of the \ps\ were reported by the Crystal
Ball~\cite{cb:cs} and BESII~\cite{bes:cs} groups, no update of its
parameters were done until 2005, when a combined fit to Crystal Ball
and BESII data was performed by Seth~\cite{seth:fit}. Recently, the
BES collaboration~\cite{bes:cs} reported new parameter values for the
$\psi(3770)$, $\psi(4040)$, $\psi(4160)$ and \ps\ resonances that are
derived from a global fit to their cross section measurements.

Despite the kinematic accessibility of ten open-charm strong decay
modes for the \ps\ its decays to any exclusive final states were not
measured until last year, when a study of exclusive \eedd\ production
via initial-state radiation (ISR) at Belle~\cite{bn989} revealed a
small enhancement near the \ps\ mass. A similar study ~\cite{prl} of
$D\overline D{}^*$ production has no evident \ps\ signal; the
$D^*\overline D{}^*$ channel exhibits a small enhancement that may be
attributable to the \ps, albeit at a slightly shifted mass value.

In this Letter we report a measurement of the exclusive cross section
for the process \eeddp\ and the first observation of \psiddt\ decay.
It represents a continuation of our studies of the exclusive open
charm production in the mass range where recently several new
charmonium-like states were observed
($Y(4260)$~\cite{babar:4260,belle:4260}, $Y(4360)$,
  $Y(4660)$~\cite{belle:4360}, $X(4160)$~\cite{belle:4160}) decaying
to either open- or closed-charm final states. Our study provides
further information on the dynamics of charm quarks at these center
of mass energies and on the properties of the \ps.  The data sample
corresponds to an integrated luminosity of $673\,\mathrm{fb}^{-1}$
collected with the Belle detector~\cite{det} at the $\Upsilon(4S)$
resonance and nearby continuum at the KEKB asymmetric-energy
  \ee\ collider~\cite{kekb}.

We select \eeddpg\ signal candidates in which the $D^0$, $D^-$ and
$\pi^+$ mesons are fully reconstructed~\cite{foot}. In general, the
\gisr\ is not required to be detected and its presence in the event is
inferred from a peak at zero in the spectrum of masses recoiling
against the \ddp\ system. The recoil mass squared is defined as:
$\RMS(\ddpf)=(\ecm - E_{\ddpf})^2 - p^2_{\ddpf}$.  Here \ecm\ is the
initial \ee\ center-of-mass (c.m.) energy, $E_{\ddpf}$ and $p_{\ddpf}$
are the c.m. energy and momentum of the \ddp\ system, respectively.
To suppress background two cases are considered: (1) the \gisr\ is out
of detector acceptance in which case the polar angle for the
\ddp\ system is required to satisfy $|\cos(\theta_{\ddp})|>0.9$; (2)
the fast \gisr\ is within the detector acceptance
($|\cos(\theta_{\ddp})|<0.9$), in the latter case the \gisr\ is
required to be detected with the mass of the $\ddp\gisr$ system
greater than $\ecm -0.58\gevc$. To suppress backgrounds from $\ee \to
D \overline D \text{n} \pi \gisr ~(\text{n}>1)$ processes, we exclude
events that contain additional charged tracks that are not used in
$D^0$, $D^-$ or $\pi^+$ reconstruction.

All charged tracks are required to originate from the vicinity of the
interaction point (IP); we impose the requirements $dr<2 \,
{\mathrm{cm}}$ and $|dz|<4\,{\mathrm{cm}}$, where $dr$ and $|dz|$ are
the impact parameters perpendicular to and along the beam direction
with respect to the IP. Charged kaons are required to have a ratio of
particle identification likelihoods, $\mathcal{P}_K = \mathcal{L}_K /
(\mathcal{L}_K + \mathcal{L}_\pi)$~\cite{nim}, larger than 0.6. No
identification requirements are applied for pion candidates. $K^0_S$
candidates are reconstructed from $\pi^+\pi^-$ pairs with an invariant
mass within $10\mevc$ of the nominal $K^0_S$ mass. The distance
between the two pion tracks at the $K^0_S$ vertex must be less than
$1\,\mathrm{cm}$, the transverse flight distance from the interaction
point is required to be greater than $0.1\,\mathrm{cm}$, and the angle
between the $K^0_S$ momentum direction and the flight direction in the
$x-y$ plane should be smaller than $0.1\,\mathrm{rad}$. Photons are
reconstructed in the electromagnetic calorimeter as showers with
energies greater than $50 \mev$ that are not associated with charged
tracks.  Pairs of photons are combined to form $\pi^0$ candidates. If
the mass of a $\gamma \gamma$ pair lies within $15\mevc$ of the
nominal $\pi^0$ mass, the pair is fit with a $\pi^0$ mass constraint
and considered as a $\pi^0$ candidate.  $D^0$ candidates are
reconstructed using five decay modes: $K^- \pi^+$, $K^- K^+$, $K^-
\pi^- \pi^+ \pi^+$, $K^0_S \pi^+\pi^-$ and $K^- \pi^+ \pi^0$.  A $\pm
15\mevc$ mass window is used for all modes except for $K^- \pi^- \pi^+
\pi^+$, where a $\pm 10\mevc$ requirement is applied ($\sim
2.5\,\sigma$ in each case). $D^+$ candidates are reconstructed using
the decay modes $K^0_S\pi^+$ and $K^- \pi^+ \pi^+$. A $\pm 15\mevc$
mass window is used for all $D^+$ modes. To improve the momentum
resolution of $D$ meson candidates, final tracks are fitted to a
common vertex with a mass constraint on the nominal $D^0$ or $D^+$
mass.  Candidates in the $D$ sideband region are selected for the
background study and refitted to the central mass value of either side
of the sideband that has the same width as the signal window.  To
remove contributions from $\ee\to D^- D{}^{*+}\gisr$,
\dnpip\ combinations with invariant mass within $\pm 10\mevc$ of the
nominal $D^{*+}$ mass are vetoed.

\begin{figure}[htb]
\begin{tabular}{cc}
\hspace*{-0.025\textwidth}
\includegraphics[width=0.47\textwidth]{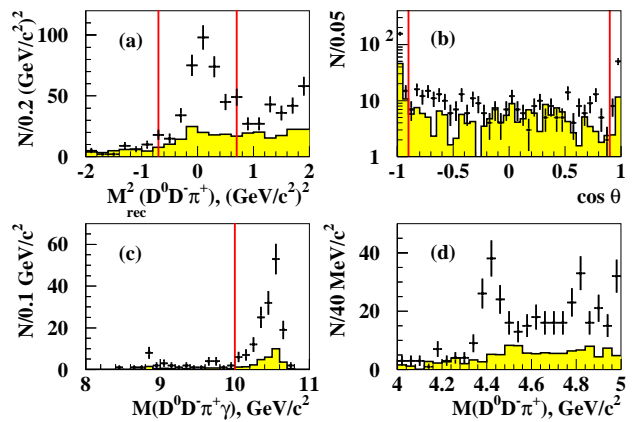}
\end{tabular}
\caption{The observed distributions of (a) $\RMS(\ddp)$ and (b)
  \ddp\ polar angles.  (c) The $\ddp\gisr$ mass spectrum. (d) The
  resulting \mddp\ spectrum.  Histograms show the normalized
  contributions from the \mdn\ and \mdm\ sidebands. The selected
  signal windows are indicated by vertical lines.}
\label{fig1}
\end{figure}

The distribution of $\RMS(\ddp)$ for the signal region is shown in
Fig.~\ref{fig1}~(a). A clear peak corresponding to the process
\eeddpg\ is evident around zero. The shoulder at higher masses is due
to $\ee \to D^0 D^- \pi^+\text{(n)}\pi^0\gisr$ events, which include
$D^{*0(-)}\to D^{0(-)}\pi^0$ decays. To suppress the tail of such
events we define the $\RMS(\ddp)$ signal region by a tight requirement
$\pm 0.7(\gevc)^2$ around zero. The polar angle distributions of
\ddp\ and the mass spectrum of the $\ddp\gisr$ combinations (when
\gisr\ is detected) in the data shown in Fig.~\ref{fig1}~(b) and
Fig.~\ref{fig1}~(c), respectively, are typical of ISR production and
agree with the Monte Carlo (MC) simulation. The \mddp\ spectrum after
the application of all requirements is shown in Fig.~\ref{fig1}~(d). A
clear peak is evident around the \ps\ mass.

The following sources of background are considered: (1) combinatorial
background under the $D^0$($D^-$) peak combined with a real
$D^-$($D^0$) coming from the signal or other processes; (2) both $D^0$
and $D^-$ candidates are combinatorial; (3) reflection from the
processes \eeddppg, with an extra $\pi^0_{\mathrm{miss}}$ in the final
state, including $D^{*0(-)}\to D^{0(-)}\pi^0_{\mathrm{miss}}$ decays;
(4) reflection from the process \eeddpgg, followed by $D^{*0}\to
D^0\gamma$, with an extra soft $\gamma$ in the final state; (5)
\misid\ in which an energetic $\pi^0$ is misidentified as a single
\gisr. The contribution of background (1) is extracted using \mdn\ and
\mdm\ sidebands that are four times as large as the signal
region. These sidebands are shifted by $30\mevc$ ($20\mevc$ for the
$D^0\to K^-\pi^-\pi^+\pi^+$ mode) from the signal region to avoid
signal over-subtraction. Background (2) is present in both the
\mdn\ and \mdm\ sidebands and is, thus, subtracted twice. To account
for this over-subtraction we use a 2-dimensional sideband region,
where events are selected from both the \mdn\ and the \mdm\ sidebands.
Backgrounds (1--2), $\sim 15\%$ of the signal, are subtracted from the
signal-region \ddp\ mass spectrum.  The dominant component of
backgrounds (3--4) is suppressed by the tight requirement on
$\RMS(\ddp)$. The remaining background from (3) is estimated by
applying a similar full reconstruction method to the
isospin-conjugated processes \eeison\ and \eeisoch. Here the absence
of additional charged tracks in the event is not required. Because of
the charge imbalance in the \ddnbp\ and \dpdmp\ final states, only
events with a missing extra $\pi^-_{\mathrm{miss}}$ can contribute to
the $\RMS(\ddnbp)$ and $\RMS(\dpdmp)$ signal windows. To extract the
level of background (3), the \ddnbp\ and \dpdmp\ mass spectra are
rescaled according to the ratio of $D^-$ and $D^0$ reconstruction
efficiencies and a factor of 1/2 due to isospin. The averaged rescaled
\ddnbp\ and \dpdmp\ mass spectrum, which is small ($\sim 2\%$ of the
signal), is subtracted from the signal-region \mddp\ spectrum.
Background (4) is also estimated from the data assuming isospin
symmetry. We measure the process $\ee\to\dstmdnp\gisr$ ($D^{*-}\to
\overline D{}^0\pi^-$) applying a similar full reconstruction
method. The shape and normalization of this spectrum after efficiency
correction is then used to generate a MC sample of $\ee \to D^{*0} D^-
\pi^+ \gisr$ ($D^{*0}\to D^0\gamma$) events. From this MC sample,
tuned to the data, the contribution from background (4) is estimated
to be less than $\sim 1.5$ events/bin, which is subtracted from the
signal \mddp\ spectrum.  The contribution from background (5) is
determined from reconstructed \misid\ events in the data found to be
negligibly small and taken into account in the systematic error.

\begin{figure}[htb]
\begin{tabular}{cc}
\hspace*{-0.025\textwidth}
\includegraphics[width=0.47\textwidth]{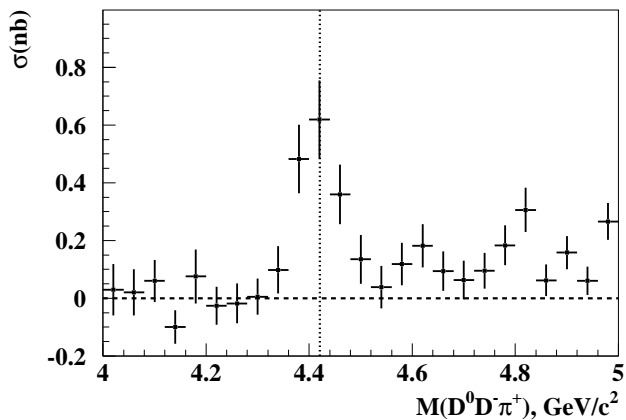} &
\end{tabular}
\caption{The exclusive cross section for \eeddp.  The dotted line
corresponds to the nominal mass of the \ps\ ~\cite{pdg}.}
\label{fig2}
\end{figure}

The \eeddp\ cross section is extracted from the background-subtracted
\ddp\ mass distribution following the procedure described
in~\cite{prl}, taking into account the differential ISR luminosity and
the total efficiency, which is found to have linear dependence on
\mddp.  The resulting \eeddp\ exclusive cross section is shown in
Fig.~\ref{fig2}. Since the bin width is much larger than resolution,
no correction for resolution is applied.

The systematic errors for the $\sigma(\eeddp)$ measurements are
summarized in Table ~\ref{tab1}.
\begin{table}[htb]
\caption{Contributions to the systematic error on the cross sections,
  [$\%$].}
\label{tab1}
\begin{center}
\begin{tabular}{@{\hspace{0.4cm}}l@{\hspace{0.4cm}}||@{\hspace{0.4cm}}c@{\hspace{0.4cm}}}
\hline \hline
Source & \ddp\
\\ \hline
Background subtraction    & $\pm 5$ \\
Cross section calculation & $\pm 5$\\
$\mathcal{B}(D)$          & $\pm 4$ \\
Reconstruction            & $\pm 6$ \\
Kaon identification       & $\pm 2$ \\
\hline
Total & $\pm 10$ \\
 \hline \hline
\end{tabular}
\end{center}
\end{table}
The systematic errors associated with the background (1--2)
subtraction are estimated to be 2\% due to an uncertainty in the
scaling factors for the sideband subtractions. It is estimated using
fits to the \mdn\ and \mdm\ distributions in the data with different
signal and background parameterization. Backgrounds (3)--(5) are also
subtracted using the data and only the uncertainty in the scaling
factors for the subtracted distribution is taken into account.
Uncertainties in backgrounds (3)--(5) are estimated conservatively to
be smaller than 4.5\% of the signal.  The systematic error ascribed to
the cross section calculation includes a 1.5\% error on the
differential luminosity and 4.5\% error in the total efficiency fit.
Another source of systematic error comes from uncertainties in track
and photon reconstruction efficiencies (1\% per track and 1.5\% per
photon). Other contributions come from the uncertainty in the kaon
identification efficiency and the absolute $D^0$ and $D^-$ branching
fractions~\cite{pdg}.

\begin{figure}[htb]
\begin{tabular}{cc}
\hspace*{-0.025\textwidth}
\includegraphics[width=0.47\textwidth]{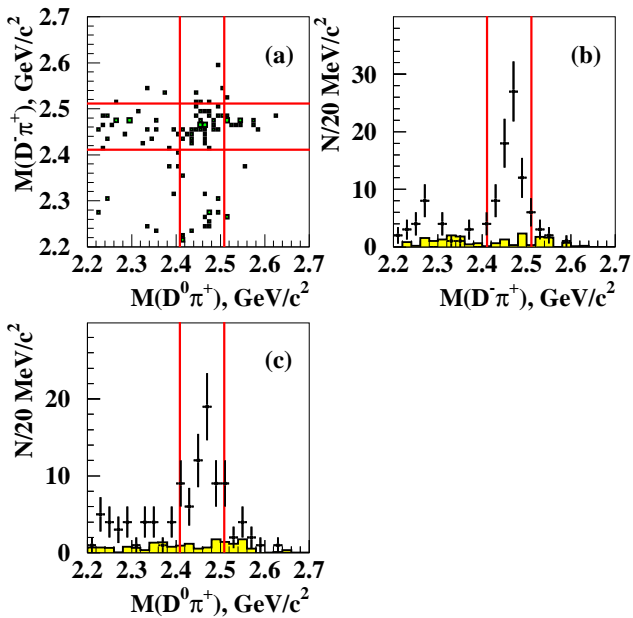}
\end{tabular}
\caption{(a) The scatter plot of \dmpip\ vs. \dnpip\ for the signal
  region in the data for $|\mddp-m_{\ps}|<100\mevc$.  (b)\dmpip\ and
  (c) \dnpip\ mass projections. Histograms show the normalized
  contributions from the \mdn\ and \mdm\ sidebands. The selected mass
  windows are illustrated by vertical and horizontal lines.}
\label{fig3}
\end{figure}

To study the resonant structure in \ps\ decays, we select
\ddp\ combinations from a $\pm 100\mevc$ mass window around the
nominal \ps\ mass ($m_{\ps}=4.421\gevc$~\cite{pdg}). A scatter plot of
$M(\dmpip)$ vs. $M(\dnpip)$ and its projections onto both axes are
shown in Figs.~\ref{fig3}~(a),(b) and (c), respectively. Clear signals
for the \dtnb\ and \dtch\ mesons are visible in these plots.  We
expect positive interference between the neutral \ddtnb\ and the
charged \ddtch\ decay amplitudes leading to the same \ddp\ final state
for the decay of $C=-1$ state, and the scatter plot evidently agrees
with this expectation.  Because of the interference we do not study
\ddtnb\ and \ddtch\ final states separately and define the signal
interval for the \ddt\ combinations as $|M_{\dmpip} - m_{\dtnb}|<50
\mevc$ or $|M_{\dnpip} - m_{\dtch}|<50 \mevc$ ($m_{\dtnb}=2.461\gevc$,
$m_{\dtch}=2.459\gevc$~\cite{pdg}).

\begin{figure}[htb]
\begin{tabular}{cc}
\hspace*{-0.025\textwidth}
\includegraphics[width=0.47\textwidth]{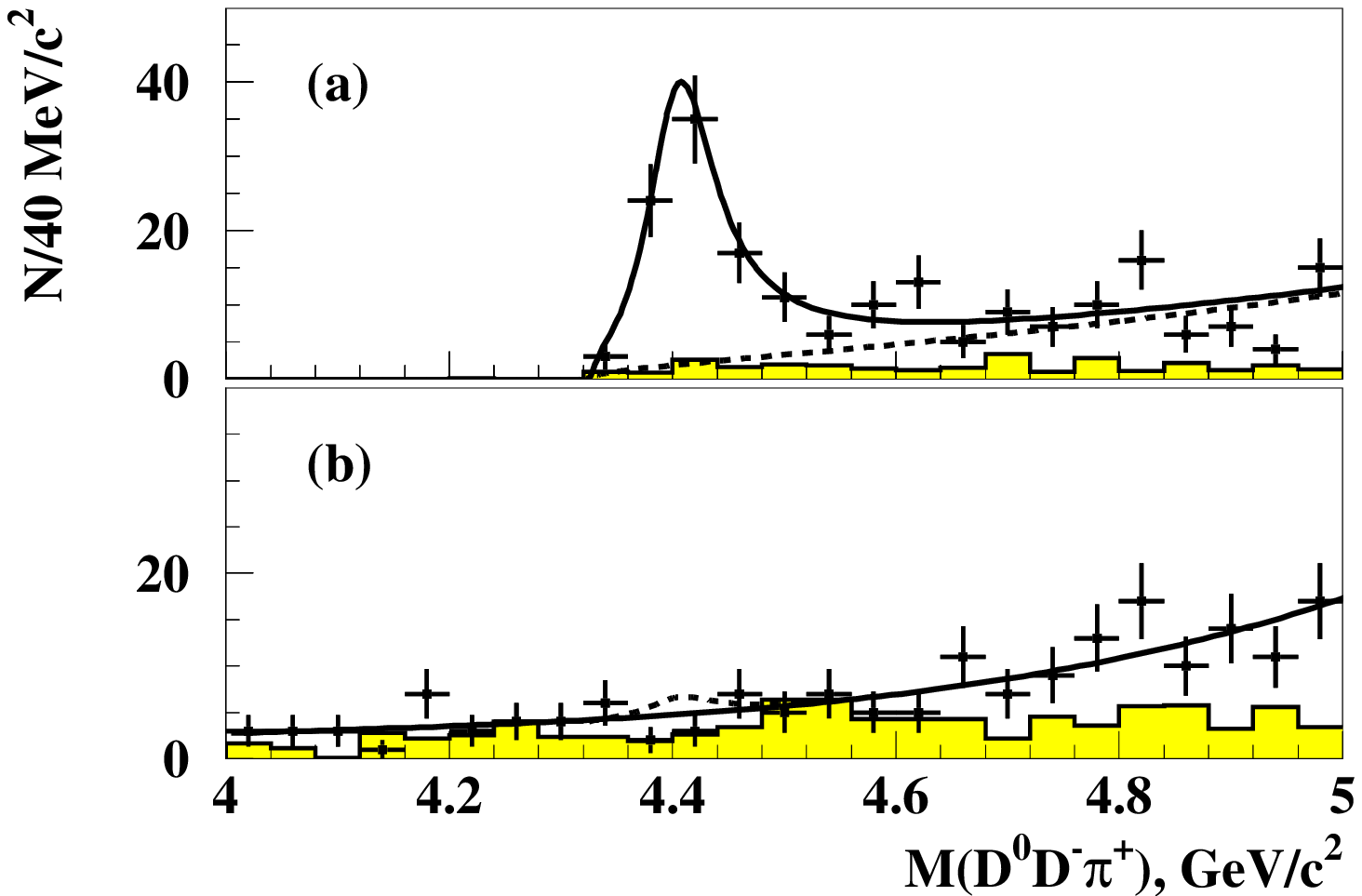}
\end{tabular}
\caption{(a) The \mddp\ spectrum for the \ddt\ signal region. The
  solid curve represents the result of the fit described in the
  text. The threshold function is shown by the dashed curve. (b) The
  \mddp\ spectrum outside the \ddt\ signal region.  The solid curve is
  a fit with a second-order polynomial background function.  The
  dashed curve shows the upper limit on the \ps\ yield at 90\% C.L.
  Histograms show the normalized contributions from \mdn\ and
  \mdm\ sidebands.}
\label{fig4}
\end{figure}

We perform a separate study of $\ee\to\ddt$ and $\ee\to D
(\overline{D} \pi)_{\text{non}\dt}$. The \mddp\ spectrum for the
\ddt\ signal interval is shown in Fig.~\ref{fig4}(a). A clear peak
corresponding to \psiddt\ decay is evident near the \ddt\ threshold.
Assuming $J^{P}=2^{+}$ for \dt\ mesons~\cite{pdg}, the signal of
$\ps\to\ddt$ should be described by a $d$-wave relativistic
Breit-Wigner (RBW) function.  However, to compare mass and width of
the obtained \ps\ signal with the corresponding \ps\ resonance
parameters measured in the {\it inclusive} study~\cite{pdg}, we
perform a likelihood fit to \mddp\ distribution with the \ddt\ signal
parameterized by an $s$-wave RBW function. The fit
  with the $d$-wave RBW function, which would be appropriate for the
  studied decay channel, is found to be unstable with respect to
  variations in the background parameterization. To account for
background and a possible non-resonant \ddp\ contribution we use a
threshold function $\sqrt{M-m_{D}-m_{\dt}}$ with a floating
normalization. Finally, the sum of the signal and background
functions is multiplied by the mass-dependent linear efficiency
function and differential ISR luminosity. The fit, shown as a solid
curve in Fig.~\ref{fig4}(a), yields $109 \pm 25\mathrm{(stat.)}$
signal events. The significance for the signal is obtained from the
  quantity $-2\ln(\mathcal{L}_0 / \mathcal{L}_{\text{max}})$, where
  $\mathcal{L}_{\text{max}}$ is the maximum likelihood returned by the
  fit, and $\mathcal{L}_0$ is the likelihood with the amplitude of the
  Breit-Wigner function set to zero.  Taking the reduction in the
  number of degrees of freedom into account, we determine the
  significance of the \ps\ signal to be $\sim10\sigma$. The obtained
peak mass $m_{\ps}=(4.411 \pm 0.007\mathrm{(stat.)})\gevc$ and total
width $\Gamma_{\mathrm{tot}}=(77 \pm 20\mathrm{(stat.)})\mevc$ are in
good agreement with the PDG~\cite{pdg} values, the recent BES
results~\cite{bes:fit} and predictions of Ref.~\cite{barnes}.
The peak cross section for $\ee \to \ps \to \ddt$ process at
$\ecm=m_{\ps}$ is calculated from the amplitude of the RBW function in
the fit to be $\sigma(\ee\to\ps) \times \mathcal{B}(\ps\to\ddt) \times
\mathcal{B}(\dt\to\dpi) = (0.74\pm 0.17 \pm 0.08)$\,nb.  Here the
systematic uncertainty is obtained by varying the fit range, histogram
bin, parameterization of the background function and efficiency. Using
$\sigma(\ee\to\ps)=12\pi/{m^2_{\ps}}\times(\Gamma_{ee}/\Gamma_{\mathrm{tot}})$
we calculate the $\mathcal{B}(\ps\to\ddt)\times
\mathcal{B}(\dt\to\dpi) = (10.5 \pm 2.4 \pm 3.8)\%$ using the
\ps\ parameters from the PDG~\cite{pdg} and $(19.5 \pm 4.5 \pm 9.2)\%$
for the \ps\ parameters from Ref.~\cite{bes:fit}.

The shape of the \mddp\ spectrum with the \ddt\ signal excluded,
shown as a solid curve in Fig.~\ref{fig4}(b), is relatively
featureless and, in the \ps\ mass window, is consistent with the
combinatorial background. The fit to this distribution with the \ps\
signal function (an $s$-wave RBW function with mass and width fixed
to the result of the previous fit) and a second-order polynomial
function yields a negative \ps\ signal that is consistent with zero.
We calculate the upper limit on the \ps\ yield to be 18 events at
90\% C.L. We assume a phase-space-like $\ps\to\ddp$ decay and
calculate an upper limit on the ratio of the branching fractions of
\ps\ decays to non-resonant \ddp\ and \ddt. This assumption is not
exactly correct since at least one meson pair has to be in a relative
$p$-wave state. However, it provides a conservative estimate of the
efficiency of the \ddt\ veto. In this way we obtain
$\mathcal{B}(\ps\to\ddp_{\mathrm{non-resonant}})/\mathcal{B}(\ps\to\ddt\to\ddp)<0.22$
at 90\% C.L.

In summary, we report first measurements of the \eeddp\ exclusive
cross section over the center-of-mass energy range from 4.0\gev\ to
5.0\gev\ with initial state radiation. Aside from a prominent
$\psi(4415)$ peak, the c.m. energy dependence of the
  \eeddp\ cross section has no evident structure. From a study of
the resonant structure in \ps\ decay we conclude that the $\ps\to\ddp$
process is dominated by \psiddt. The mass and width of the \ps\ state
are found to be $(4.411\pm 0.007\mathrm{(stat.)})\gevc$ and $(77\pm
20\mathrm{(stat.)})\mevc$, respectively.

We thank the KEKB group for excellent operation of the accelerator,
the KEK cryogenics group for efficient solenoid operations, and the
KEK computer group and the NII for valuable computing and Super-SINET
network support.  We acknowledge support from MEXT and JSPS (Japan);
ARC and DEST (Australia); NSFC and KIP of CAS (China); DST (India);
MOEHRD, KOSEF and KRF (Korea); KBN (Poland); MES and RFAAE (Russia);
ARRS (Slovenia); SNSF (Switzerland); NSC and MOE (Taiwan); and DOE
(USA).


\begin{thebibliography} {99}

\bibitem{mark:cs} J.~Siegrist {\it et al.} (Mark-1 Collab.),
Phys. Rev. Lett. {\bf 36}, 700 (1976).

\bibitem{dasp:cs} R.~Brandelik {\it et al.} (DASP Collab.),
Phys. Lett. B {\bf 76}, 361 (1978).

\bibitem{seth:fit} K.~K.~Seth, Phys. Rev. D {\bf 72}, 017501 (2005).

\bibitem{cb:cs} A.~Osterheld {\it et al.} (Crystal Ball Collab.),
SLAC-PUB-4160, 1986.

\bibitem{bes:cs} J.~Z.~Bai {\it et al.} (BES Collab.),
Phys. Rev. Lett. {\bf 88}, 101802 (2002).

\bibitem{bes:fit} M.~Ablikim {\it et al.} (BES Collab.),
arXiv:0705.4500 [hep-ex].

\bibitem{bn989} G.~Pakhlova, arXiv:0708.0082 [hep-ex] (2007).

\bibitem{prl} G.~Pakhlova {\it et al.} (Belle Collab.),
  Phys. Rev. Lett. {\bf 98}, 092001 (2007).

\bibitem{babar:4260} B.~Aubert {\it et al.} (BaBar Collab.),
  Phys. Rev. Lett. {\bf 95}, 142001 (2005).

\bibitem{belle:4260} C.~Z.~Yuan {\it et al.} (Belle Collab.),
  arXiv:0707.2541 [hep-ex](2007), accepted by Phys. Rev. Lett.

\bibitem{belle:4360} X.~L.~Wang {\it et al.} (Belle Collab.),
  arXiv:0707.3699 [hep-ex] (2007), accepted by Phys. Rev. Lett.

\bibitem{belle:4160} K.~Abe {\it et al.} (Belle Collab.),
 arXiv: 0708.3812 [hep-ex](2007), submitted to Phys. Rev. Lett.

\bibitem{det} A.~Abashian {\it et al.} (Belle Collab.),
  Nucl. Instr. and Meth.  A {\bf 479}, 117 (2002).

\bibitem{kekb} S.~Kurokawa and E.~Kikutani, Nucl. Instr. and Meth. A
  {\bf 499}, 1 (2003); and other papers included in this volume.

\bibitem{foot} Charge-conjugate modes are included throughout this
  paper.

\bibitem{nim} E.~Nakano, Nucl. Instr. and Meth. A {\bf 494}, 402
(2002).

\bibitem{pdg} W.-M.~Yao {\it et al.} (Particle Data Group), J. Phys.
G {\bf 33}, 1 (2006).

\bibitem{barnes} T.~Barnes, S.~Godfrey and E.~S.~Swanson,
Phys. Rev. D {\bf 72}, 054026 (2005).

\end{thebibliography}
\end{document}